# Interferometry from Space: A Great Dream

Erik Høg
*Niels Bohr Institute, Juliane Maries Vej 30, 2100 Copenhagen Ø, Denmark*

During some thirty years, 1980-2010, technical studies of optical interferometry from instruments in space were pursued as promising for higher spatial resolution and for higher astrometric accuracy. Nulling interferometry was studied for both high spatial resolution and high contrast. These studies were great dreams deserving further historical attention. ESA's interest in interferometry began in the early 1980s. The studies of optical interferometry for the global astrometry mission GAIA began in 1993 and ended in 1998 when interferometry was dropped as unsuited for the purpose, and the Gaia mission to be launched in 2013 is not based on interferometry. © Anita Publications. All rights reserved.

## Introduction

In retrospect, the great vision of optical interferometry from space, 1980-2010, may be called a dream because all technical studies have been stopped and only one optical mission with interferometry has been realized, viz in the Hubble Fine Guidance Sensors. The dream may however become reality once when technological and economic conditions permit since the scientific purposes remain highly valid.

ESA's interest in interferometry began in the early 1980s, according to Michael Perryman, see the appendix. The studies of optical interferometry for the global astrometry mission GAIA began in 1993 and ended in 1998 when interferometry was dropped. The Gaia mission to be launched in 2013 is not based on interferometry, but on direct imaging on CCDs by large telescopes. It is well known that optical interferometry on the ground and radio interferometry on the ground and from space have been successful, but this is not considered here.

The present note is based on a proposal of September 2011 for a session in C41 on the subject with coorganizers C. Beichman and M. Fridlund. We listed six proposals for interferometry, labeled a-f.

## A list of proposals a-f

a. Proposals for high spatial resolution as well as high contrast were the European DARWIN – ended in 2007, see [1] and [2], and the NASA Terrestrial Planet Finder (TPF-I).

b. Proposals for astrometry were: The US series of proposals: POINTS, SONATA – small OSI, SIM PlanetQuest [3], SIM Lite - finally cancelled in 2010, see [4], and the StarLight mission.

c. Astrometry with the Hubble Fine Guidance Sensors has been successful, see [5].

d. A series of US proposals for astrometry: NEWCOMB, FAME-1, but the interferometry was dropped in 2003, see [6].

e. The European astrometric mission GAIA95 with interferometry and its small version DIVA were abandoned in 2004, see [6].

f. The European proposal for astrometry "GAIA" in 1993, where the "I" stands for interferometry, led to the astrometric Gaia mission due for launch in 2013, but without interferometry, see [6] and [7].

## GAIA and Gaia

In October 1993, ESA received proposals for Cornerstone mission studies. The proposers for astrometry asked for study of "a large Roemer option and an interferometric option", GAIA. They should be studied as two concepts for an ESA Cornerstone Mission for astrometry "without a priori excluding

*Corresponding author :*
*e-mail: Erik.hoeg@get2net.dk* (Prof Erik Høg)

either". The Roemer mission based on non-interferometric imaging had been proposed the year before, in September 1992 and this had started the activity in the Hipparcos community towards a new astrometric mission.

In March 1997, ESA invited members to a science advisory group on astrometry at the 10 microarcsec level by space interferometry. The ESA letter [7] (see quote in [6] p.10) primarily selects interferometry for study: "Space interferometry was identified in the ESA long-term programme for space science, Horizon 2000, as a potential candidate among space projects planned for after the turn of the century…" and then the letter mentions astrometry as the first option for study. Most probably, without the interferometry in the GAIA proposal, astrometry would not have been selected for study at all.

The letter does not mention the proposers' request to study also a non-interferometric option, and it should be emphasized that the astrometrists had already been working on astrometry by interferometry believing that was the way to go. But the study by industry quickly showed that this was wrong. Interferometry was abandoned already in January 1998 and the further development was "a large Roemer option" with direct imaging on CCDs and detection by Time-Delayed Integration (TDI). The GAIA mission without interferometry was approved by ESA in 2000, and the name was changed to Gaia about 2004.

I have traced the early years of Gaia development in a report [6], including the various interferometric designs proposed for astrometry. It would be interesting to trace the decision processes in favour of interferometry in working groups and responsible committees. In case of astrometry, the alternative option without interferometry was ignored by ESA in 1997, against the request from the proposing scientists to study both options.

In conclusion, it is noted that we worked seriously during 1993-97 to develop space astrometry by interferometry, according to ESA's vision. The mission has kept its name from this failing design concept but we easily accepted the simpler imaging and detection technique for Gaia proposed by industry in 1998, which was already contained in the Roemer proposal of 1992. The Roemer proposal also started the work on a successor to Hipparcos and ESA's director of science Alvaro Gimenez recently wrote to me: "I am of course well aware of the essential role of the Roemer proposal for GAIA to become a reality today."

## Acknowledgements

I am grateful to Charles Beichman and Malcolm Fridlund for comments and information. I am indebted to Michael Perryman for his recent information about how GAIA was born, and to Virginia Trimble for comments to an earlier version of this paper.

**APPENDIX**

The appendix contains (1) a recent account by Michael Perryman about ESA's interest in interferometry since the early 1980s, and on how GAIA was born in the evening of 9 September 1993, (2) highlights from the development of Gaia, and (3) an extensive list of references on space astrometry in the 1990s.

## 1 Michael Perryman on how GAIA was born

On 4 November 2013 Michael Perryman, ESA project scientist for Gaia 1995-2007, replied to my question about ESA's interest in interferometry with the following which I quote with his kind permission:

Indeed, I knew about ESA's interest in interferometry, and that knowledge came to astrometry mainly through me and Lennart. Here is the relevant extract from my own notes: "Meanwhile, and in fact since the early 1980s, there had been ESA study teams devoted to space interferometry. Under Sergio Volonté's direction, I had coordinated some of the work around 1988, leading to a report on a 'strategy' for space interferometry. Although interferometry had appeared in the Horizon 2000 planning as a 'Green Dream', and therefore to some extent inevitable, I was unhappy with its immediate prospects - low throughput, a relatively small number of accessible scientific targets, complexity, UV coverage, time variability - all seemed to undermine its scientific applicability. Interferometry was also being looked at from the lunar perspective, but this too, I felt, was somewhat of a waste of time: it will come to nothing in my lifetime, and I do not think that the moon is as obvious a platform for this type of work as general hype would have us believe (astrometry benefits from all-sky visibility, for example). I sat in on one or two of the lunar meetings, with my (external) colleagues Lennart Lindegren, Jean Kovalevsky, Jan Noordam, Alan Greenaway, and Sergio Volonte (Paris). ESA was going through preparations to start the Horizon 2000+ long-term plan. Proposals were being solicited. What should we do?

On 9 September 1993, there was a lunar interferometry meeting in ESTEC. My diary entry "Useful in that it jogged me into putting a proposal in to Post Horizon 2000 planning, for a project which I called GAIA, for Global Astrometric Interferometer for Astrophysics". In the evening, I met up with Lennart Lindegren, and we had dinner together at the Camino Real, Leiden, and we discussed Hipparcos, and possibilities for the future. This is when Gaia was born - I believed that we could exploit the present goodwill and interest in interferometry with a proposal for Horizon 2000+. The acronym was coined, by me, during the day. That evening, Lennart and I agreed to prepare and submit a proposal. A letter was duly prepared and submitted to ESA by Lennart; we prepared an outline proposal, added the names of interested people, and Lennart submitted it on 12 October 1993."

## 2 Highlights from the development of Gaia

The Figs 1 and 2 show the Roemer and GAIA proposals from respectively 1992 and 1993/94. Figure 3 shows the first design of GAIA by the industrial contractor in 1997, a system with three interferometers which was soon abandoned and replaced by direct imaging on CCDs. These three figures are adapted from [6] where more explanation is given. The final design of Gaia as implemented in 2011 appears in the figs 4-7.

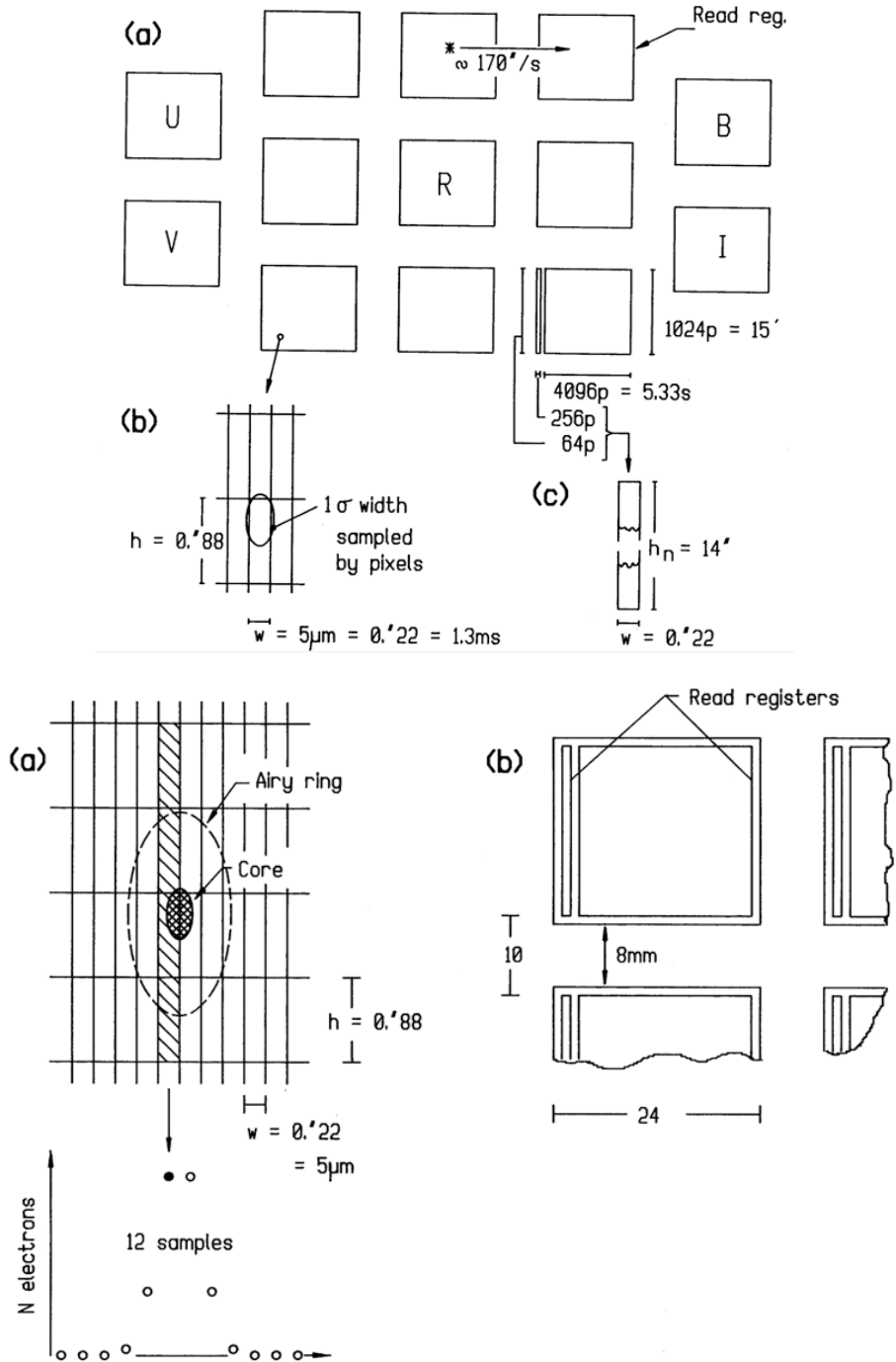

Fig. 1. Left: Focal plane of the Roemer satellite with CCDs in scanning mode proposed in 1992, stars moving left to right through the field (Høg 1993). Right: Sampling of the Roemer CCDs. (a): The pixels are elongated perpendicular to the scanning, several pixels (here 4) are read together in order to decrease the readnoise, and (b): short CCDs are provided for bright stars, features later adopted in Gaia.

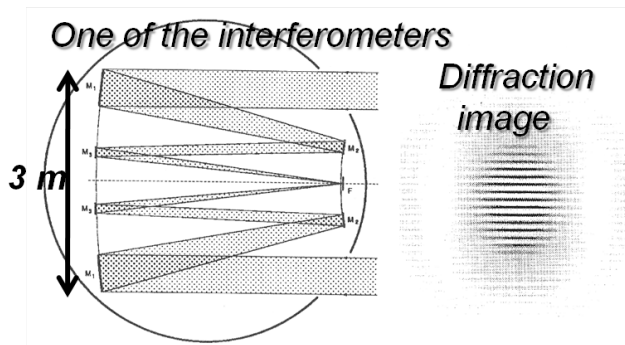

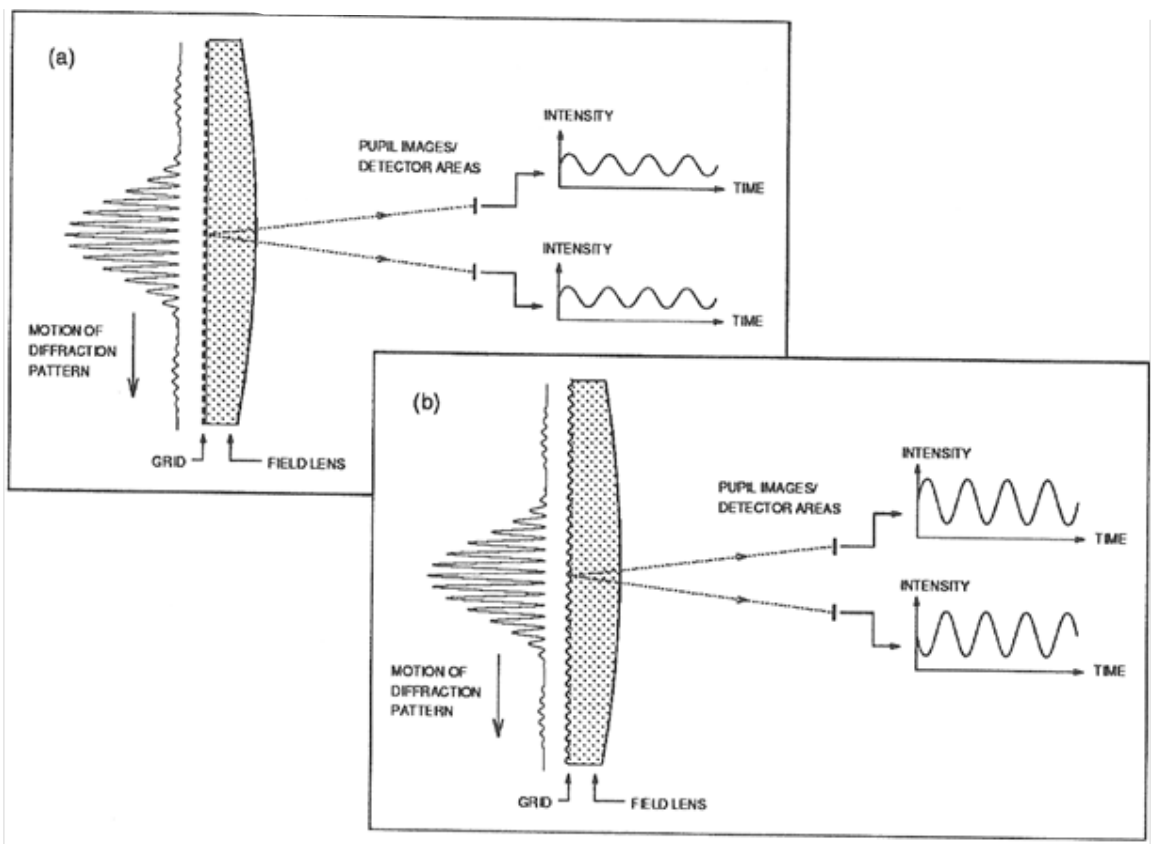

Fig. 2. The GAIA system proposal as it appeared in August 1994 (Lindegren & Perryman 1995). Left: The optical system. Right: detection of the modulation.

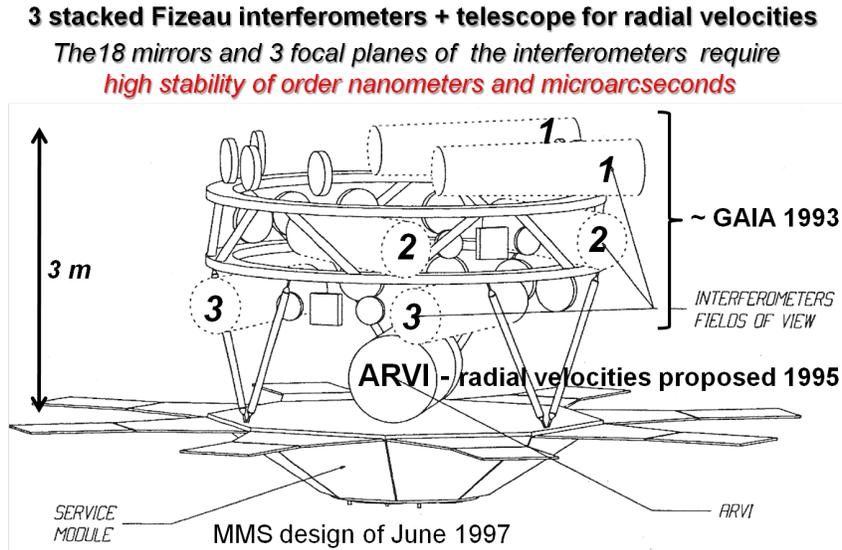

Fig. 3. The GAIA design of June 1997 from the study proposal by MMS (Matra Marconi Space). The detection system with a modulating grid, shown in Fig. 2, was dropped by MMS a few months later and by ESA in January 1998 and only detection of the diffraction image of the stars directly on the CCD with full telescope aperture was further considered.

## Payload and Telescope

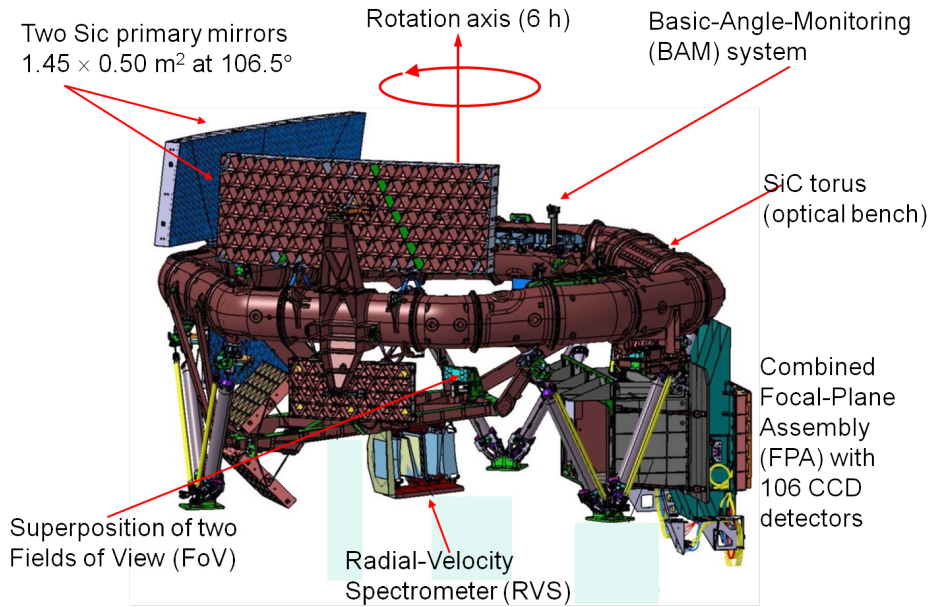

Fig. 4. The Gaia payload and telescope in 2011. Courtesy by EADS-Astrium – 2011.

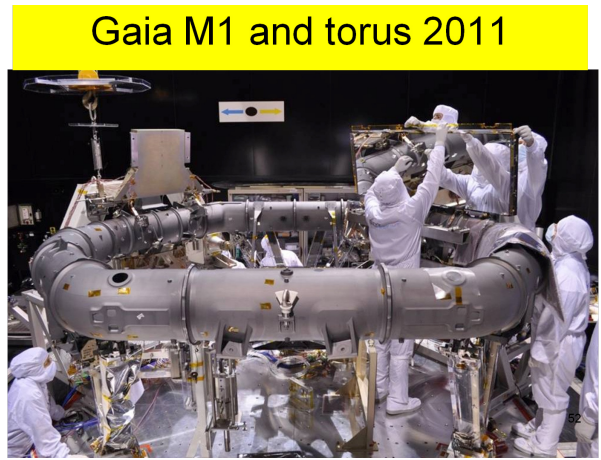

Fig. 5. The Gaia torus with one of the mirrors being mounted. Courtesy by EADS-Astrium – 2011.

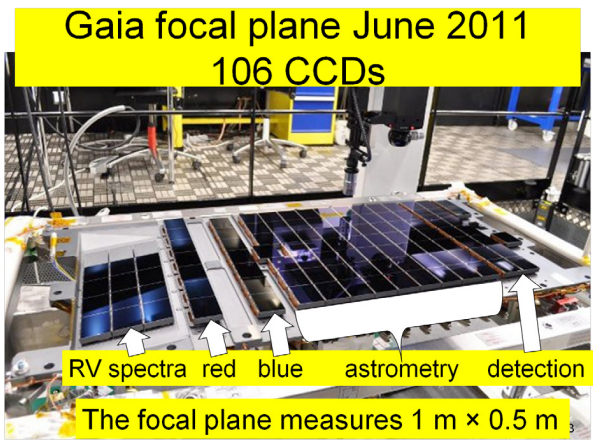

Fig. 6. The Gaia focal plane. Courtesy by EADS-Astrium – 2011.

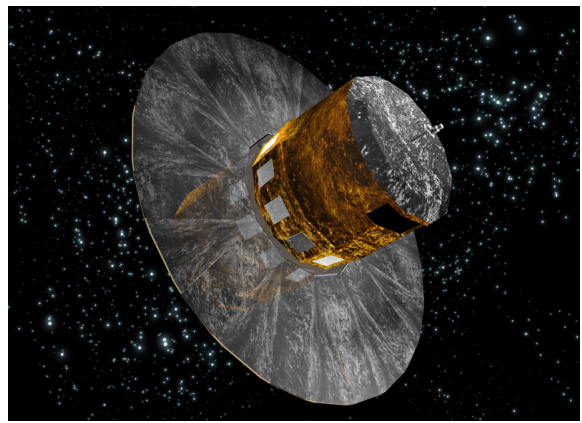

Fig. 7. The Gaia satellite with the large sun shield as it will appear in orbit. Courtesy by EADS-Astrium – 2011.

**3 References on space astrometry**

The following list was compiled for the report [6] on the development of space astrometry especially in the 1990s, but more references are included than mentioned in the report though of course without being complete.

The references are given in the sequence they were presented publicly at conferences or by other distribution. The year of publication in proceedings is often a year later. Reports on Hipparcos and Tycho at the symposia are mostly omitted.

**1980:**

Di Serego Alighieri S, Rodgers A W, Stapinski T E 1980, An alternative detector for Hipparcos. Padova – Asiago Obs. / M.S.S.S.O. Technical Report No. 2, January 1980. As received from M. Perryman in 2011, available at www.astro.ku.dk/~erik/AltDetHip.pdf

**Leningrad in October 1989:**

Chubey M S, Makarov V V, Yershov V N, Kanayev I I, Fomin V A, Streletsky Yu S, Schumacher A V 1990, A Proposal for Astrometric Observations from Space. In: J H Lieske and V K Abalakin (eds.) *Inertial Coordinate Systems on the Sky* Leningrad, USSR, Oct 17-21, 1989, IAU Symp No 141, 77.

Chandler J F and Reasenberg R D 1990, **POINTS:** A Global Reference Frame Opportunity. In: J H Lieske and V K Abalakin (eds.) *Inertial Coordinate Systems on the Sky.* Leningrad, USSR, Oct 17-21, 1989, IAU Symp No 141, 217.

Duncombe R L, Jefferys W H, Benedict G F, Hemenway P D, Shelus P J 1990, Expectations for Astrometry with the Hubble Space Telescope In: J H Lieske and V K Abalakin (eds.) *Inertial Coordinate Systems on the Sky* Leningrad, USSR, Oct 17-21, 1989, IAU Symp No 141, 339.

Seidelmann P K 1990, Space Astrometry and the HST Wide Field/Planetary Camera. In: J H Lieske and V K Abalakin (eds.) *Inertial Coordinate Systems on the Sky* Leningrad, USSR, Oct 17-21, 1989, IAU Symp No 141, 347.

Nesterov V V, Ovchinnikov A A, Cherespachuk A M, Sheffer E K 1990, The **LOMONOSOV** Project for Space Astrometry. In: J H Lieske and V K Abalakin (eds.) *Inertial Coordinate Systems on the Sky* Leningrad, USSR, Oct 17-21, 1989, IAU Symp No 141, 355.

Avanesov G A, Vavaev V A, Ziman Ya L *et al* 1990, **REGATTA-ASTRO** Project: Astrometric studies from Small Space Laboratory. In: J H Lieske and V K Abalakin (eds.) *Inertial Coordinate Systems on the Sky* Leningrad, USSR, Oct. 17-21, 1989, IAU Symp No 141, 361.

Stone R C and Monet D G 1990, The USNO (Flagstaff Station) CCD Transit Telescope and Star Positions Measured from Extragalactic Sources. In: J H Lieske and V K Abalakin (eds.) *Inertial Coordinate Systems on the Sky.* Leningrad, USSR, Oct 17-21, 1989, IAU Symp No 141, 369.

**Moscow in June 1991:**

Høg E, Chubey M 1991. Proposal for a second Hipparcos. Presented at the International Symposium "Etalon" Satellites held in June 1991 in Moscow and as poster at the IAU General Assembly in Buenos Aires in August. It was not published but is now scanned and the 6 pages are available at www.astro.ku.dk/~erik/ Hipparcos-2.pdf

**Shanghai in September 1992:**

Høg E and Lindegren L 1993, A CCD Modulation Detector for a Second Hipparcos Mission. In: I I Mueller and B Kolaczek (eds.) *Developments in Astrometry and Their Impact on Astrophysics and Geodynamics* Shanghai, China, Sept 15-19, 1992, IAU Symp No 156, 31. At http://esoads.eso.org/ abs/1993IAUS..156...31H

Høg E 1993, Astrometry and Photometry of 400 Million Stars Brighter than 18 Mag. In: I.I. Mueller and B Kolaczek (eds.) *Developments in Astrometry and Their Impact on Astrophysics and*

*Geodynamics* Shanghai, China, Sept. 15-19, 1992, IAU Symp No 156, 37. Available at http://esoads.eso.org/abs/1993IAUS..156...37H

**Proposal for ESA M3 mission in May 1993:**

Lindegren L (ed), Bastian U, Gilmore G, Halbwachs J L, Høg E, Knude J, Kovalevsky J, Labeyrie A, van Leeuwen F, Pel J W, Schrijver H, Stabell R, Thejll P 1993a. ROEMER: Proposal for the Third Medium Size ESA Mission (M3). Lund Observatory.

**Cambridge in June 1993:**

Morrison and G F Gilmore (eds.) 1994, *Galactic and Solar System Optical Astrometry.* Proceedings of a conference held in Cambridge June 21-24, 1993. 19+339 pp.

Høg E and Lindegren L 1994, ROEMER satellite project: the first high-accuracy survey of faint stars. In: L V Morrison and G F Gilmore (eds.) *Galactic and Solar System Optical Astrometry.* Proceedings of a conference held in Cambridge June 21-24, 1993. 246-252.

**To ESA in October 1993:**

Lindegren L, Perryman MAC, Bastian U, Dainty J C, Høg E, van Leeuwen F, Kovalevsky J, Labeyrie A, Mignard F, Noordam J E, Le Poole R S, Thejll P, Vakili F 1993b. GAIA: Global Astrometric Interferometer for Astrophysics, proposal for a Cornerstone Mission concept submitted to ESA on 12 October 1993, including a cover letter by Lindegren, 6 pages. Available at www.astro.ku.dk/~erik/gaia_proposal_1993.pdf

**March 10, 1994:**

Seidelmann writes to me that USNO is pursuing NEWCOMB which is however unfunded, and that they are interested in collaboration on other space-based astrometry projects like Roemer and GAIA.

**April 1994:**

Lindegren L 1994. Letter in April 1994 to Steven Beckwith, chairman of the UV-to-radio topical team of the Survey Committee then drafting ESA's Horizon 2000+ plan, 1+7 pages. Available at www.astro.ku.dk/~erik/beckwith_lindegren.pdf

**The Hague in August 1994:**

Høg E, Seidelmann P K (eds) 1995, *Astronomical and Astrophysical Objectives of Sub-milliarcsecond Optical Astrometry,* Proceedings of the 166[th] Symposium of the IAU, held in The Hague, The Netherlands, August 15-19, 1994,16+441 pp.

Høg E 1995a, A New Era of Global Astrometry. II: A 10 Microarcsecond Mission. (Including proposal of Roemer+). In: Proceedings of the 166[th] Symposium of the IAU, 317.

Johnston K, Seidelmann P K, Reasenberg R D, Babcock R, Philips J D 1995, Newcomb Astrometric Satellite. In: Proceedings of the 166[th] Symposium of the IAU, 331.

Lindegren L and Perryman MAC 1995, A Small Interferometer in Space for Global Astrometry: The GAIA Concept. In: Proc. of the 166[th] Symposium of the IAU, 337.

**To ESA in September 1994:**

Lindegren L, Perryman MAC 1994. GAIA: Global Astrometric Interferometer for Astrophysics (A Concept for an ESA Cornerstone Mission). Supplementary Information Submitted to the Horizon 2000+ Survey Committee.

**Cambridge in June 1995:**

Perryman MAC, van Leeuwen F, eds 1995, Proceedings of a Joint RGO-ESO Workshop on "Future Possibilities for Astrometry in Space", Cambridge, UK, 19-21 June 1995 (**ESA SP-379**, September 1995), 323 pp.

Some of the following reports are available at www.astro.ku.dk/~erik/papers/

**FAME-1 in 1995:**

**GAIA95 report in November 1995:**

Interferometer for Astrometry and Spectrophotometry", both versions were widely distributed. It appeared in Experimental Astronomy in 1997.

Høg E, Fabricius C, Makarov V V 1997, GAIA95: Astrometry from Space: New Design of the GAIA Mission. Experimental Astronomy 7: 101.

**1996:**

Röser S, Bastian U, de Boer K S, Høg E, Schalinski C, Schilbach E, de Vegt Ch, Wagner S 1996, August. DIVA: Deutsches Interferometer für Vielkanalphotometrie und Astrometrie. Antrag für die Phase A Studie.

Seidelmann P K 1998, Prospects for Future Astrometric Missions. In: B J McLean *et al* (eds.), *New Horizons from Multi-Wavelength Sky Surveys,* Proceedings of the 179[th] Symposium of the IAU, held in Baltimore, USA, August 26-30, 1996, 79.

**1997:**

Cecconi M, Gai M, Lattanzi M G 1997, A New Optical Configuration for GAIA. In: ESA SP-402, 803.

Cecconi M, Rigoni G, Bernacca P L 1997, Opto/Mechanical Analysis for the Space Interferometry Project GAIA. In: ESA SP-402, 811.

ESA 1997, Invitation to join the ESA Science Advisory Group on Astrometry Cornerstone Study. Available at www.astro.ku.dk/~erik/esa_sag_invitation_970311.pdf

Gai M, Bertinetto F, Bisi M, *et al* 1997, GAIA Feasibility: Current Research on Critical Aspects. In: ESA SP-402, 835.

Høg E, Bastian U, Seifert W 1997, Optical Design for GAIA. In: ESA SP-402, 783.

Lindegren L, Perryman MAC 1997, GAIA: Global Astrometric Interferometer for Astrophysics. In: ESA SP-402, 799.

Perryman MAC, Bernacca PL (eds.) 1997, HIPPARCOS Venice '97, Presentation of the Hipparcos and Tycho Catalogues. ESA SP-402, 52+862 pp.

Reasenberg RD, Babcock RW, Chandler J F, Phillips J D 1997, POINTS Mission Studies: Lessons for SIM. Center for Astrophysics Preprint Series No. 4496 (Received February 28, 1997). Invited paper for a conference at STEcI in October 1996. 15 pages. It contains e.g. some instrument descriptions of POINTS and NEWCOMB but no designs.

Scholz R D,  Bastian U 1997, Simulated Dispersed Fringes of an Astrometric Space Interferometry Mission. In: ESA SP-402, 815.

**Gotha in May 1998:**

Lindegren L 1998, Hipparcos and the future: GAIA. In: P Brosche *et al* (eds) Proceedings of the International Spring Meeting of the Astronomische Gesellschaft, Gotha, May 11-15, 1998, "The Message of the Angles – Astrometry from 1798 to 1998", Acta Historica Astronomiae, vol. 3, 214. http://esoads.eso.org/abs/1998AcHA....3..214L

Høg E, Fabricius C, Knude J, Makarov V V 1998, Sky survey and photometry by the GAIA satellite. In: P Brosche *et al* (eds) Proceedings of the International Spring Meeting of the Astronomische Gesellschaft, Gotha, May 11-15, 1998, "The Message of the Angles – Astrometry from 1798 to 1998", Acta Historica Astronomiae, vol 3, 223.

http://esoads.eso.org/abs/1998AcHA....3..223H

**Leiden in November 1998:**

Leiden 1998, Proceedings of the GAIA Workshop, November 23-27, 1998 in Leiden. In: Baltic Astronomy, An international journal, Vol 8, Nrs 1 and 2, 324 pp, 1999.

**2000:**

ESA 2000, GAIA, Composition, Formation and Evolution of the Galaxy, Concept and Technology Study Report. ESA-SCI(2000)4, 381pp. Available in "Library & Livelink" at

www.rssd.esa.int/index.php?project=GAIA&page=index

**Les Houches in May 2001:**

Les Houches 2001, GAIA: A European Space Project – A summer school 14-18 May 2001. O Bienayme, C Turon (eds.), EAS Publication Series, Volume 2, 395 pp, 2002.

**Vilnius in July 2001:**

Vilnius 2001, Census of the Galaxy: Challenges for Photometry and Spectrometry with GAIA. Vansevicius V, Kucinskas A, Sudzius J (eds.), Proceedings of the Workshop held in Vilnius, Lithuania, 2-6 July 2001. From: Astrophysics and Space Sciences Vol. 280, Nos. 1-2, 194 pp, 2002.

**2003:**

Johnston K J 2003, The FAME Mission. In: Proceedings of the SPIE, Volume 4854, 303. At 2003SPIE.4854...303J

**2007:**

Høg E 2007, From the Roemer mission to Gaia. A poster at IAU Symposium No. 248 in Shanghai, October 2007. Only the first three pages appear in the Proceedings. Complete version available at www. astro.ku.dk/~erik/ShanghaiPoster.pdf and as report No.4 at www.astro.ku.dk/~erik/History.pdf

**2008:**

Høg E 2008a, Bengt Strömgren and modern astrometry: Development of photoelectric astrometry including the Hipparcos mission. At www.astro.ku.dk/~erik/History.pdf and as poster at IAU Symposium No. 254.

Høg E 2008b, Astrometric accuracy during 2000 years. www.astro.ku.dk/~erik/Accuracy.pdf

Høg E 2008c, Astrometry and Optics during the Past 2000 Years, a collection of 9 reports. At www. astro.ku.dk/~erik/History.pdf.

Høg E 2008d, Four lectures about the General History of Astrometry. At www.astro.ku.dk/~erik/ Lectures.pdf

**2009:**

Gouda N *et al* 2009, Series of JASMINE projects ---Exploration of the Galactic bulge --- http:// www.ast.cam.ac.uk/iau_comm8/iau27/presentations/Gouda.pdf

Hennessy G S, Gaume R 2009, Space astrometry with the Joint Milliarcsecond Astrometry Pathfinder. Proc. of the IAU Symposium 261, 350.

**2011:**

Høg E 2011a, Miraculous approval of Hipparcos in 1980: (2). At www.astro.ku.dk/~erik/ HipApproval.pdf

Høg E 2011b, Astrometry Lost and Regained. Or: From a modest experiment in Copenhagen in 1925 to the Hipparcos and Gaia space missions. At www.astro.ku.dk/~erik/AstromRega3.pdf

Høg E 2011c, Lectures on Astrometry. www.astro.ku.dk/~erik/Lectures2.pdf

Jasmine 2011, European-Japanese collaboration. Nano-JASMINE and AGIS

www.rssd.esa.int/index.php?project=GAIA&page=picture_of_the_week&pow=132

SIM 2011, http://en.wikipedia.org/wiki/Space_Interferometry_Mission